\documentclass[12pt,preprint]{aastex}
\usepackage{epsfig}
\usepackage{natbib}
\usepackage{lscape}
\makeatletter
\newcommand*{\rom}[1]{\expandafter\@slowromancap\romannumeral #1@}
\makeatother
  \usepackage{hyperref}
\usepackage{amsmath} 

  \usepackage{courier}
\usepackage{graphicx}
 
\shorttitle{Hot Companion of 60 Cygni}
\shortauthors{Wang et al.}

\begin{document}

\received{}
\accepted{}
\setcounter{footnote}{0}

\title{Detection of the Ultraviolet Spectrum \\ of the Hot Subdwarf Companion 
 of 60 Cygni (B1~Ve) \\ from a Survey of IUE Spectra of Be Stars}

\author{Luqian Wang, Douglas R. Gies} 
\affil{Center for High Angular Resolution Astronomy and  
 Department of Physics and Astronomy,\\ 
 Georgia State University, P. O. Box 5060, Atlanta, GA 30302-5060, USA} 
\email{lwang@chara.gsu.edu, gies@chara.gsu.edu} 
 
\author{Geraldine J. Peters\altaffilmark{1}} 
\affil{Space Sciences Center, University of Southern California, Los Angeles, CA 90089-1341, USA}  
\email{gpeters@usc.edu} 
 
\altaffiltext{1}{Guest Observer with the {\it International Ultraviolet Explorer}.} 
 
\slugcomment{Accepted for ApJ, 5/12/2017} 
\paperid{AAS05063}

\begin{abstract}
We used archival {\it International Ultraviolet Explorer} {(\it IUE)} 
high-dispersion, short wavelength spectra data to search for 
evidence of the spectra of hot subdwarf companions of 
six rapidly rotating Be stars in binary systems. 
We searched for the signature of a hot companion through an analysis of the 
cross-correlation functions of observed and model spectra that were separated
into primary and secondary components using a Doppler tomography algorithm 
and adopted spectroscopic orbital solutions.  A positive detection of the flux 
from a hot companion was made for the reconstructed secondary cross-correlation function 
of just one target, 60~Cygni (B1 \rom{5}e).  We estimate that the companion of the Be star in 60~Cygni 
has $T_{\rm eff} = 42 \pm 4$ kK, mass ratio $M_2/M_1 = 0.15 \pm 0.02$, and monochromatic 
flux ratio ${f_2}/{f_1} = 0.034 \pm 0.002$ in the spectral region 
near 1525 \AA.  If the companions of the other target Be stars are also hot, 
then they must be faint and contribute less than $\approx 1\%$ of the UV flux  
($<0.6 \%$ in the case of $\gamma$~Cas).
We also discuss in an appendix a shell episode of Pleione (28~Tau) recorded in 
the {\it IUE} spectra.    
\end{abstract}

 
\setcounter{footnote}{1}


\section{Introduction} 

Be stars are B-type, main-sequence stars showing hydrogen emission features in their 
optical spectra \citep{rivinius2013}.   The Balmer line and excess continuum emission 
originates from a circumstellar disk that is inherently variable on time scales
from days to centuries.  Be stars are generally fast rotating objects, with equatorial 
rotating speeds of $>70\%$ of their critical velocities.   Their disks result from 
processes that are shedding the stellar angular momentum, and 
disk formation episodes may be linked to pulsational mode beating in some Be stars (e.g., $\eta$~Cen and $\mu$~Cen; \citealt{baade2016}).

The cause of the rapid rotation of Be stars is due in some cases to past 
mass transfer in interacting binary systems \citep{pols1991,shao2014}.
Roche lobe overflow of the initially more massive, mass donor star will
lead to spin-up of the gainer and an increase in orbital period once 
mass ratio reversal occurs.  The final stripped-down donor star may explode 
to create a Be star -- neutron star X-ray binary, or, if the remnant is 
below the Chandrasekhar mass limit, the binary system will consist of a Be star 
with a faint low-mass white dwarf or helium subdwarf star (sdO).
Detection of such subdwarf companions is difficult because they are relatively 
faint and their low mass creates only small orbital motions in the Be star 
companions. 

Nevertheless, we have now detected the flux of subdwarf companions in four systems. 
The search is best made at short wavelengths because the subdwarfs are usually 
hotter than their Be star companions and thus contribute relatively more flux 
in the ultraviolet (UV).  The first direct spectroscopic detection of a sdO companion 
in the Be system $\phi$~Per was accomplished with UV spectra from the 
{\it International Ultraviolet Explorer (IUE)} \citep{thaller1995}, and the 
binary properties were established through {\it Hubble Space Telescope} UV 
spectroscopy \citep{gies1998} and optical interferometry \citep{mourard2015}.  
Subsequently, \citet{peters2008} and \citet{peters2013} applied similar search techniques
with {\it IUE} to detect subdwarf companions in the Be binaries FY~CMa and 59~Cyg, respectively.
Finally, the flux of a relatively very faint companion was detected by \citet{peters2016}
from a large set of {\it IUE} spectra of HR~2142.  The gravitational pull of the 
companion creates a tidal wake in the disk of the Be star in HR~2142 that leads to the 
appearance of ``shell lines'' (narrow absorption features) when the subdwarf is in the foreground. 
These four Be+sdO binaries may be related to recently discovered systems consisting 
of an intermediate mass, main-sequence star and a hot, low mass white dwarf that was stripped 
of its envelope by binary mass transfer \citep{maxted2014,matson2015,rappaport2015}. We note 
for completeness that there are three known late B-type plus white dwarf systems (detected 
through their flux in the extreme ultraviolet; \citealt{vennes1997,burleigh1998,burleigh1999}), 
but these are thought to be non-interacting systems in which the white dwarf formed by
a single-star evolutionary route. 

Based on the positive detection of companions in these four Be systems, here we expand the 
search for subdwarfs in another six Be binary systems that were observed many times over 
the lifetime of {\it IUE}.  All of these targets are known single-lined spectroscopic 
binaries with hitherto undetected, low mass companions.   The targets are listed in 
Table~1 with spectral classifications from \citet{slettebak1982}
and projected rotational velocities from \citet{fremat2005}.  
We briefly summarize below the key studies on their orbital and physical properties. 
\newline
\textbf{$\gamma$ Cas}  {(\bf{HD 5394})}. This is the first spectroscopically observed 
Be star by \citet{secchi1866}. The binarity of the system was confirmed by 
\citet{nemravova2012} through a 16.84 year span of optical spectra monitoring.  
Extensive studies of the star's X-ray emission and circumstellar 
environment \citep{smith2012,stee2012,hamaguchi2016} suggest that the X-ray emission 
is associated with parts of
the disk gas of the Be star, so that the companion need not be an accreting 
neutron star or white dwarf.  Several other X-ray emitting Be stars like $\gamma$~Cas appear to be 
``blue stragglers'' that may have been formed through binary mass transfer \citep{smith2016}. 
\newline
\textbf{28 Tau} {(\bf{HD 23862})}. Pleione is a known shell star with cyclic variability lasting decades. 
The shell activity that occurred over the duration of the {\it IUE} observations is discussed in the Appendix.  
\citet{nemravova2010} confirmed it as the primary component of the binary system through 
Doppler shifts from H$\alpha$ emission spectroscopy. 
\newline
\textbf{$\zeta$ Tau} {(\bf{HD 37202})}. \citet{struve1942} confirmed the binarity of the system using 
radial velocities from Balmer lines measured in photographic plates, and \citet{delplace1971} 
estimated the system dimensions by considering the size of the Be disk. 
\citet{ruzdjak2009} improved the orbital ephemeris of the system from H$\alpha$ spectroscopy 
and $UBV$ photometry observations over about a century. 
\newline
\textbf{$\kappa$ Dra} {(\bf{HD 109387})}. \citet{saad2005} determined the orbit of the binary through wide-ranging 
spectroscopic observations.  They found no direct evidence of the companion's spectrum.   
\newline
\textbf{60 Cyg} {(\bf{HD 200310})}. \citet{plaskett1931} reported that the star belongs 
to a spectroscopic binary system. \citet{koubsky2000} discussed the long, medium, and
short term variability in the observed spectra and light curves, and they presented 
orbital elements from radial velocity measurements of the Be star.    
\newline
\textbf{$\pi$ Aqr} {(\bf{HD 212571})}. \citet{bjorkman2002} discovered the binary nature through a 
radial velocity analysis of the absorption and emission lines.  They suggested 
that the companion could be an A- or F-type main-sequence star, but the absorption
spectrum of the companion was not observed.  \citet{zharikov2013} made an analysis
of the orbital variations of the H$\alpha$ emission line.  They estimated 
that the secondary has a mass $<2 M_\odot$. 

\begin{deluxetable}{ccccccc}
\tablenum{1}
\tablecaption{\emph{IUE} Observations of Be Binary Systems}
\tablehead{
\colhead{Star } & 
\colhead{HD} & 
\colhead{HIP} & 
\colhead{Spectral} & 
\colhead{$V\sin i$} & 
\colhead{Number of}  & 
\colhead{Time span}   \\
\colhead{Name} & 
\colhead{Number} & 
\colhead{Number} & 
\colhead{Classification} & 
\colhead{(km s$^{-1}$)} & 
\colhead{Observations} & 
\colhead{(years)}}
\startdata
$\gamma$ Cas & \phn\phn5394 & \phn\phn4427 & B0.5 IVe       & 432 &    228 & 18 \\
28 Tau       &    \phn23862 &    \phn17851 & B8 Ve shell    & 286 & \phn48 & 17 \\
$\zeta$ Tau  &    \phn37202 &    \phn26451 & B1 IVe shell   & 310 & \phn34 & 17 \\
$\kappa$ Dra &       109387 &    \phn61281 & B5 IIIe        & 200 & \phn26 & 11 \\
60 Cyg       &       200310 &       103732 & B1 Ve          & 300 & \phn23 & 15 \\
$\pi$ Aqr    &       212571 &       110672 & B1 III-IVe     & 230 & \phn22 & 17
\enddata
\tablecomments{Spectral classifications and projected rotational velocities are from 
\citet{slettebak1982} and \citet{fremat2005}, respectively.}
\end{deluxetable}

Here we present results of our analysis of the UV spectra from \emph{IUE} 
to search for subdwarf companions of these six systems.  
Section 2 presents our subdwarf flux search method that is based upon 
a cross-correlation analysis of the UV spectra with model spectral templates 
for the subdwarf spectra.  Our results are discussed in Section 3.  


\section{Search for the UV Flux of Hot Companions}

Our search technique is centered on forming a cross-correlation function 
(CCF) of the observed spectra with model spectra for the expected effective 
temperature $T_{\rm eff}$ of a hot companion.  Two issues make this 
process difficult.  First, any companion is probably relatively faint
compared to its Be host star, so given the low S/N of the {\it IUE} spectra 
($\approx 20$ per resolution element; \citealt{nichols1996}),
it will be hard to identify the signal of the
companion in the CCF for any one spectrum.  Second, some of the 
lines in the model spectrum of the hot companion may also be 
present (although weaker) in the spectrum of the Be star, so the 
derived CCF may present a composite of companion and Be signals. 
Our solution here is to apply a Doppler tomography algorithm 
\citep{bagnuolo1994} to the derived CCFs in order to separate both 
the companion and Be star CCF components and to use all the spectra 
together to increase the net S/N of the reconstructed CCFs. 
Below we outline how the spectra are organized, how the trial velocity 
curves were estimated for the components, and details of the CCF 
reconstruction process.  The final results for the reconstructed 
secondary CCFs are discussed at the end of the section and are shown 
in Figure~1.  

We downloaded the high dispersion, short wavelength prime (SWP) spectra 
of the selected targets from MAST\footnote{https://archive.stsci.edu/iue/}.
The number of available spectra and their time span are listed in Table 1.
The spectra have a resolving power of $\lambda/\Delta\lambda = 13,000$, 
and the wavelength coverage is 
from 1150 $\mbox{\AA}$ to 1950 $\mbox{\AA}$.  The individual echelle orders 
were merged and placed on a heliocentric grid using the standard \emph{IUE} 
IDL data reduction pipeline {\tt iuerdaf}.   The spectra were then 
transformed onto a uniform heliocentric wavelength grid in $\log \lambda$ 
steps equivalent to 10 km~s$^{-1}$, and for the more distant stars 
(HD 5394, HD 200310, and HD 212571 with $d > 180$ pc), the wavelength 
calibration was checked through alignment (and then removal) of interstellar
lines.  The spectra were rectified with respect to the relatively line-free 
parts of the spectra into a final matrix of stellar flux as a function of
wavelength and time of observation. 

The next aspect of the process was to estimate the radial velocities of 
both components at the time of observation.  We initially measured radial 
velocities of the Be components using the CCF methods adopted in past work
\citep{peters2016}, but the scatter in the results (and poor phase coverage
in some cases) was too large to offer any improvement over the existing 
orbital solutions.  Consequently we adopted the published orbital elements 
for the Be star primaries that are summarized in Table 2.   Then we calculated 
velocities for the primary Be star at the times of observation using 
these solutions.  The secondary star velocities were derived from the 
primary velocities using the systemic velocity $\gamma$ and an 
assumed mass ratio $q=M_2/M_1$.   This mass ratio was calculated from 
the spectroscopic mass function and published estimates of the Be star 
mass and system inclination.  The latter two parameters were taken from 
\citet{silaj2014} and the references cited in Table~2. 
The derived values of $q$ are given in column 8 of Table~2.   


\begin{deluxetable}{cccccccccc}
\rotate
\tabletypesize{\scriptsize}
\tablenum{2}
\tablecaption{Adopted Orbital Elements}
\tablewidth{0pt}
\tablehead{
\colhead{HD} & 
\colhead{$P$} & 
\colhead{$T$\tablenotemark{a}} & 
\colhead{$e$} & 
\colhead{$\omega$} & 
\colhead{$K_1$} &
\colhead{$\gamma$} & 
\colhead{Trial} &  
\colhead{$f_2/f_1$} &
\colhead{Source} \\
\colhead{Number}  & 
\colhead{(days)} & 
\colhead{(HJD-2,400,000)} & 
\colhead{} & 
\colhead{(deg)} & 
\colhead{(km\ s$^{-1}$)} & 
\colhead{(km\ s$^{-1}$)} & 
\colhead{$M_2/M_1$} & 
\colhead{} &
\colhead{Reference} }
\startdata
\phn\phn5394 & $203.523 \pm 0.076$ & $52183.65 \pm 0.62$ & 0.0               & \nodata         & $4.30 \pm 0.09$ & $0.02 \pm 0.06$ & 0.072 & $<0.006$ & \citet{nemravova2012} \\
   \phn23862 & $218.023 \pm 0.023$ & $40040.4  \pm 1.6$  & $0.596 \pm 0.035$ & $147.7 \pm 4.5$ & $5.41 \pm 0.35$ & $-0.15\pm 0.1$  & 0.082 & $<0.026$ & \citet{nemravova2010} \\
   \phn37202 & $132.987 \pm 0.050$ & $47025.6  \pm 1.8$  & 0.0               & \nodata         & $7.43 \pm 0.46$ & $20.0 \pm 0.4$  & 0.084 & $<0.013$ & \citet{ruzdjak2009} \\
      109387 & $ 61.555 \pm 0.029$ & $49980.22 \pm 0.59$ & 0.0               & \nodata         & $6.81 \pm 0.24$ & \nodata         & 0.114 & $<0.010$ & \citet{saad2005}   \\
      200310 & $146.6   \pm 0.6$   & $50016.9  \pm 1.9$  & 0.0               & \nodata         & $10.8 \pm 0.1$  & $-13.4$         & 0.131 & $ 0.034$ & \citet{koubsky2000}  \\
      212571 & $84.135  \pm 0.004$ & $50316.91 \pm 0.04$ & 0.0               & \nodata         & $16.7 \pm 0.2$  & $-4.9 \pm 0.1$  & 0.163 & $<0.010$ & \citet{bjorkman2002}  
      \enddata
\tablenotetext{a}{For systems with circular orbits, the epoch corresponds to the time of Be star maximum radial velocity, and 
for the eccentric orbit of HD~23682, the epoch corresponds to the time of periastron. }
\end{deluxetable}

Each spectrum was cross-correlated with a model spectrum that was derived from the grid of 
synthetic spectra for hot stars using the non-local thermal equilibrium (non-LTE) atmosphere program 
TLUSTY and the associated radiative transfer code SYNSPEC from \citet{lanz2003}.
The default model temperature was set to $T_{\rm eff}=45$~kK, a value similar to that found
for other hot subdwarf systems \citep{peters2016}, and the gravity was assigned a value of 
$\log g = 4.75$, the highest gravity available in the grid but probably lower than expected. 
The model spectrum is based upon a solar abundance and a microturbulent velocity of 10 km~s$^{-1}$.
The synthetic spectrum was rectified to a unit continuum, rebinned onto the observed $\log \lambda$ grid,
and smoothed to the instrumental broadening of FWHM = 25 km~s$^{-1}$.  No rotational broadening 
was applied given the sharp appearance of the subdwarf spectra in other Be binary systems. 
We excluded the beginning and ending regions, plus very broad wind lines from the spectra 
before calculating the CCF to avoid introduction of troublesome wide structures in the CCF.
The regions adjoining the excluded zones were gradually smoothed to a pure continuum using a Tukey 
filter function, which applies a cosine function to reduce line depths close to the boundaries. 
The final result is an array of CCF functions corresponding to each observed spectrum. 

The derived CCF functions usually display a wide and shallow peak because of some correlation between 
the broad lines of the Be star spectrum and the model spectrum.  We suppose that each CCF is the 
sum of a broad component from the Be star and a possible narrow component from the subdwarf star, 
each Doppler-shifted accord to the orbital velocity at the time of observation.  Then we 
performed a reconstruction of these two components using the Doppler tomography algorithm of 
\citet{bagnuolo1994}.  This procedure assumes radial velocity shifts from the adopted orbital 
elements (Table~2), and then uses an iterative scheme to reconstruct the two components that when 
shifted and co-added make the best fit of the observed CCFs. In this instance, we assumed that
both components contribute equally to the combined CCF, and consequently we divided the results
by a factor of two so that the CCF amplitude matches that in the observed CCFs.  The iterative 
scheme begins by assuming that the primary CCF is equal to the simple mean of all the CCFs and that
the secondary CCF is flat and featureless (no secondary signal is present).  

The reconstructed primary component appears almost flat in the cases of the cooler Be star 
primaries (28 Tau and $\kappa$ Dra) because there is very little correlation between the 
features in their and the model spectra.  However, the primary CCFs for the hotter Be stars
($\gamma$~Cas, $\zeta$~Tau, 60~Cyg, $\pi$~Aqr) are broad as expected for the broad and 
shallow absorption lines in their spectra that are common to those in the hot model spectrum.
The trial reconstructed CCFs for the secondaries are generally shallow with some low 
frequency curvature due to the characteristics of the spectral rectification and line clusters.  
In order to remove the variations in the background, we first smoothed the secondary CCF 
by convolution with a Gaussian of FWHM = 350 km~s$^{-1}$, and then we subtracted this smoothed 
background variation from the secondary CCF to search for any narrow peak associated with 
a hot subdwarf spectral component. 
The resulting secondary CCFs are plotted in Figure~1, and they 
generally appear featureless, with the striking exception of 60~Cyg, in which we see 
a prominent peak near the expected zero velocity in the frame of reference of the 
secondary star.  Thus, our search method has led to the discovery of one new detection 
of a hot companion to the Be star 60~Cyg.  In the next section, we discuss the properties 
of the companion of 60~Cyg and consider the flux limits of the companions of the other 
five targets.  

\begin{figure}[h!] 
\begin{center} 
{\includegraphics[angle=90,height=12cm]{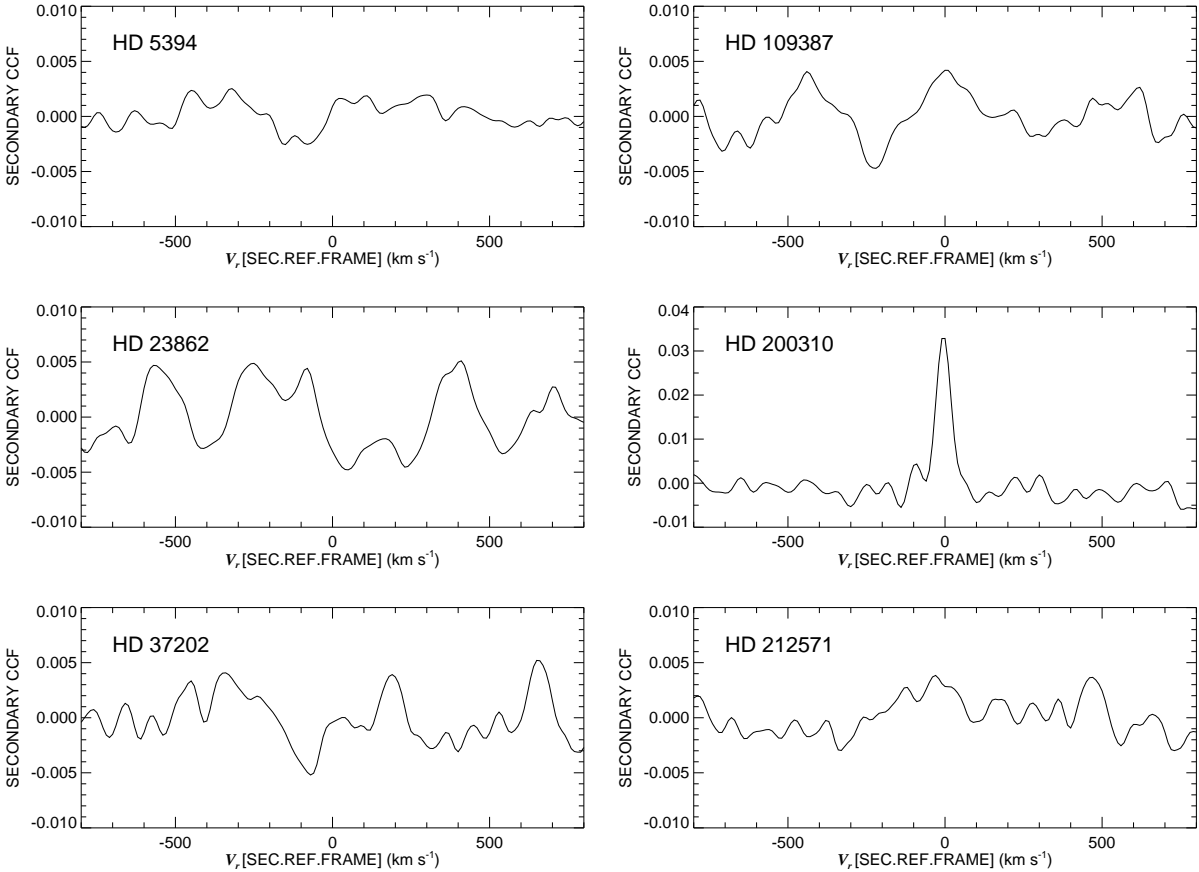}} 
\end{center} 
\caption{The reconstructed secondary CCFs of the six Be binary systems derived from the assumed 
Doppler shifts of the secondary and a model hot reference spectrum ($T_{\rm eff}=45$~kK). 
A detected peak is only seen in the case of 60~Cyg = HD~200310. \label{fig1}} 
\end{figure} 


\section{Discussion}

The detection of the spectral signature of the hot companion of 60~Cyg involved making 
assumptions about the mass ratio and companion effective temperature.  We next explore
how changing these parameters can help us find better estimates of their values. 
We began by repeating the analysis for a grid of assumed mass ratios, and in each case
we fit a Gaussian to the peak in the reconstructed secondary CCF.  The CCF peak attains 
a maximum at $q = M_2/M_1 = 0.146 \pm 0.023$, where the uncertainty represents the 
half-range of mass ratios where the peak height is within the 1 $\sigma$ zone of the
maximum CCF peak height of $0.0335 \pm 0.0015$.  We note that this estimate of the 
mass ratio is the same within uncertainties as the trial mass ratio we derived from 
the primary mass function, mass, and system inclination (Table~1). 
We fixed the mass ratio to this result, and then repeated the analysis using TLUSTY/SYNSPEC 
model spectra for the cross-correlation template that ranged from $T_{\rm eff} = 27.5$ to 55~kK.  
We made a spline fit of peak height as a function of assumed temperature to arrive at an 
estimate of $T_{\rm eff} = (42 \pm 4)$~kK for the companion. 

We are also interested in determining the UV flux ratio that would produce a secondary 
CCF with the observed peak height.  We determined this quantity by first creating 
a subdwarf model spectrum for $T_{\rm eff} = 42$~kK, $\log g =4.75$, $V \sin i = 0$, and constructing 
a Be star model for $T_{\rm eff} = 27$~kK, $\log g = 4.0$, $V \sin i = 320$ km~s$^{-1}$ \citep{koubsky2000}.  
Then we created a model set of observations over an assumed
grid of flux ratios in the following way.
For each time of observation, the model spectra were shifted by the assigned orbital 
Doppler velocities.  Next the model secondary spectrum was rescaled in flux to 
match that of the model primary spectrum in the range 1500 -- 1550 \AA .
Then the model secondary spectrum was rescaled again in flux according to
the assumed monochromatic flux ratio $f_2/f_1$.  The sum of the model primary
and rescaled secondary spectra was then rectified to a unit continuum.  
We then repeated the CCF and tomography analysis of this model set of spectra
in the same way as done with the observed spectra  
to derive a relationship between reconstructed secondary CCF peak height 
and the assumed flux ratio $f_2/f_1$ at 1525 \AA .
Interpolating the observed CCF peak height in this relation led to a monochromatic flux 
ratio estimate of $f_2/f_1 = 0.0339 \pm 0.0015$ at 1525 \AA .   
This flux ratio is related to the ratio of stellar radii by
\begin{equation}
\frac{f_2}{f_1} = \frac{F_2}{F_1}\times \Big(\frac{R_2}{R_1}\Big)^2
\end{equation}
where $F_2/F_1$ is the monochromatic surface flux ratio of the stars and 
is equal to $F_2/F_1 = 3.88$ in the continuum at 1525 \AA\ for the adopted 
temperatures of the two stars in the TLUSTY models. Using the above relation, 
the resulting radius ratio is $\frac{R_2}{R_1} = 0.093 \pm 0.012$, where the 
uncertainty includes the range of $F_2/F_1$ associated with the error in $T_{\rm eff}$.

\citet{koubsky2000} estimate that the Be star mass and mean radius of 60~Cyg are 
$M_1 = 11.8 M_\odot$ and $R_1 = 5.1 R_\odot$, respectively.  If we adopt these values, 
then the mass, radius, and gravity of the subdwarf are $M_2 = 1.7 M_\odot$, $R_2 = 0.48 R_\odot$, 
and $\log g_2 = 5.3$, respectively.  This estimate of the companion mass places it 
above the Chandrasekhar limit of $1.4 M_\odot$, so it is possible that the subdwarf 
may become a neutron star in the future.  If so, then 60~Cyg may be a progenitor of 
a Be X-ray binary system.  The estimated gravity of the subdwarf is higher than the 
limit in the TLUSTY grid $\log g = 4.75$ that we used in deriving the best fit 
$T_{\rm eff}$.  The relative strengths of the spectral features in the model spectra 
depend on the ionization balance in the model, so if we had used a higher $\log g$ 
model, we would probably arrive at a slightly higher estimate of $T_{\rm eff}$
and a somewhat lower value of $R_2$.

It is remarkable that among the now five known Be+subdwarf systems that 60~Cyg and
59~Cyg are separated by about one degree on the sky.  The stars are not known
as cluster members, but they may be affiliated with other early-type stars in
their vicinity.  We formed a list of stars with B spectral classifications
within $2^\circ$ of the midpoint between 59~Cyg and 60~Cyg from the catalog
of \citet{skiff2014}, and then collected proper motions from the \citet{gaia2016}
and parallaxes from \citet{leeuwen2007} for these stars.  Table 3 shows those
selected with proper motions and parallaxes within $2\sigma$ of those of
59 Cyg (HD 200120) and 60 Cyg (HD 200310).  This collection of six objects
may form part of a hitherto unknown co-moving group of equal age stars.  If so, 
then 59~Cyg and 60~Cyg may be at a similar stage of binary evolution because of
their common origin and age.


\begin{deluxetable}{ccccccc}
\tablenum{3}
\tablecaption{Co-moving Stars in the Vicinity of 59 Cyg and 60 Cyg}
\tablewidth{0pt}
\tablehead{
\colhead{HD} &
\colhead{$V$} &
\colhead{Spectral} &
\colhead{$\mu_\alpha$} &
\colhead{$\mu_\delta$} &
\colhead{$\pi$} &
\colhead{$V_r$} \\
\colhead{Number}  &
\colhead{(mag)} &
\colhead{Classification} &
\colhead{(mas y$^{-1}$)} &
\colhead{(mas y$^{-1}$)} &
\colhead{(mas)}  &
\colhead{(km s$^{-1}$)}  }
\startdata
198625 &  6.33  & B4 Ve       & $4.2 \pm 1.0$  &    $3.8 \pm 1.0$  &  $2.03 \pm 0.34$ & $-15$ \\
199309 &  8.65  & B8 V        & $6.4 \pm 0.6$  &    $2.6 \pm 0.6$  &  $3.08 \pm 0.70$ & $-25$ \\
199889 &  8.33  & B8 V        & $6.4 \pm 0.6$  &    $3.1 \pm 0.6$  &  $3.10 \pm 0.52$ & $-22$ \\
200120 &  4.75  & B1.5 Vnne   & $7.3 \pm 1.0$  &    $2.5 \pm 1.0$  &  $2.30 \pm 0.42$ & $-10$ \\
200310 &  5.43  & B1.5 IV-Vne & $6.1 \pm 1.0$  &    $3.4 \pm 1.0$  &  $2.14 \pm 0.37$ & $-13$ \\
200615 &  8.13  & B8 V        & $6.0 \pm 0.7$  &    $1.4 \pm 0.6$  &  $1.69 \pm 0.53$ & $-17$ 
\enddata
\tablecomments{The spectral classifications are from the collection by \citet{skiff2014}, 
proper motions $\mu_\alpha$ and $\mu_\delta$ are from the \citet{gaia2016}, and parallaxes 
are from \citet{leeuwen2007}. The $V$ magnitude and radial velocity data are from SIMBAD, 
with the exception of the systemic velocities for 59 Cyg \citep{peters2013} and 60 Cyg \citep{koubsky2000}.}
\end{deluxetable}

We found no evidence of the companion flux in the other five Be binaries. 
We placed a constraint on the relative flux of the undetected companion 
in the same way as we found the flux ratio for 60~Cyg by creating model 
observation sets for the adopted stellar parameters and then finding 
the relation between $f_2/f_1$ at 1525 \AA\ and the secondary CCF peak height. 
We used these relations and the observed standard deviations in the 
derived CCFs (Fig.~1) to estimate a conservative upper limit on $f_2/f_1$
that would yield a $5 \sigma$ peak in the CCF.  These limits on the monochromatic
flux are listed in column 9 of Table~2, and in general they indicate that
the companions contribute no more than $1\%$ of the flux at 1525 \AA .
We caution that these limits tacitly assume that the companions are 
hot and narrow-lined, which may not be correct in all cases. 
Nevertheless, the results suggest that if the companions are hot, 
then they must have small radii and relatively low luminosity.
For example, the flux ratio limit is the smallest in the case of $\gamma$~Cas, 
$f_2/f_1 < 0.006$, and this limit implies $R_2 / R_\odot < 0.3$ and 
$\log L_2 / L_\odot < 2.6$.
The faintness of such companions is probably consistent with evolutionary
models that predict that such post-mass transfer remnants spend most
of their remaining He-burning lifetime as faint, core He-burning objects
(A.\ Schootemeijer et al.\ 2017, in preparation). 


\acknowledgments

The data presented in this paper were obtained from the 
Mikulski Archive for Space Telescopes (MAST). STScI is operated by the 
Association of Universities for Research in Astronomy, Inc., under 
NASA contract NAS5-26555. Support for MAST for non-HST data is provided 
by the NASA Office of Space Science via grant NNX09AF08G and by other 
grants and contracts.
Our work was supported in part by NASA grant NNX10AD60G (GJP) and 
by the National Science Foundation under grant AST-1411654 (DRG). 
Institutional support has been provided from the GSU College
of Arts and Sciences, the Research Program Enhancement
fund of the Board of Regents of the University System of Georgia
(administered through the GSU Office of the Vice President
for Research and Economic Development), and by the USC Women in
Science and Engineering (WiSE) program (GJP). 
This work has made use of data from the European Space Agency (ESA)
mission {\it Gaia}, processed by
the {\it Gaia} Data Processing and Analysis Consortium (DPAC). Funding
for the DPAC has been provided by national institutions, in particular
the institutions participating in the {\it Gaia} Multilateral Agreement.  


Facilities: \facility{IUE}


\newpage

\appendix
\section{Shell Episode of Pleione}

Pleione shows long term variations in brightness and color during the B, Be, and Be-shell phase transitions. 
During a shell phase the optical spectrum of hydrogen and metallic lines show narrow absorption cores with 
significant growth in line strength, and these spectral variations probably result from physical changes in 
the circumstellar disk as projected against the photosphere of the star. 
\citet{cramer1995} summarized the light variations of Pleione from 1880 to 1993 through optical photometry, 
and they documented how the $B$-magnitude of Pleione changed during the shell episode from 1973 to 1993. 
A significant drop in brightness occurred prior to the shell phase in 1973, and then the star's brightness 
gradually increased as the shell lines faded and Pleione returned to the Be phase. 
\citet{doazan1988} investigated the spectral variability of the metallic lines in far-UV during this time, 
and \citet{doazan1993} showed how the energy flux between 1250 and 3000 \AA\ varied from 1979 to 1991. 

The full set of \emph{IUE} spectra record the shell line changes from 1979 to 1995, and we plot the average 
flux of Pleione in the UV from 1160 to 1450 \AA\ in Figure 2 (top panel). The average UV flux dropped 
significantly after 1979, reached a minimum in 1982, and then slowly increased back to the average level. 
The three low flux measurements (1980 $-$ 1982) were recorded through the small aperture of the camera, 
which may underestimate the actual flux. Nevertheless, a similar local minimum was also recorded at the 
same time in the $B$-band photometry from \citet{cramer1995}. There are shell lines in the UV spectra that 
are very narrow and that are also observed in cooler objects like A-type stars. Thus, we measured the 
strength of shell lines of Pleione by calculating CCFs with a star that has a line spectrum similar to the shell 
features. We adopted the supergiant star HD 197345  (Deneb, A2\rom{1}a) to cross-correlate with the 
spectrum of Pleione, and shell line regions including 1450 $-$ 1500 \AA, 1510 $-$ 1583 \AA, 1650 $-$ 1700 \AA, 
and 1828 $-$ 1875 \AA\ were chosen to calculate the CCFs. In Figure 2, the middle panel shows how the shell 
line strength decreased as the star brightened. We also calculated the relative radial velocity of the shell 
lines from the peak position of CCFs with the Deneb spectrum. \citet{parthasarathy1987} reported that Deneb has a mean 
radial velocity of $-$13.67 km~s$^{-1}$, so we added this velocity to the CCF relative velocities to obtain the 
absolute radial velocities of the shell lines that are plotted in bottom panel of Figure 2. The shell line 
velocities appear to become more negative relative to Pleione (systemic velocity of $-$0.15 km s$^{-1}$ 
from \citealt{nemravova2010}) as the shell episode declines. However, near the final stage of the shell episode, 
the CCFs are weak, so the measured velocities have large uncertainties. Our results are consistent with those 
reported by \citet{doazan1982} in their Figure 11$-$43, which show that the velocities have greatest expansion 
at the epoch of shell line disappearance. The shell line spectra show significant blue-shifts during the 
Be-shell to Be phase transition, and the negative velocities suggest that the opaque gas associated with the 
shell lines moves away from the star as the shell episode concludes.

\begin{figure}[h!] 
 \centering
 \includegraphics[height=13cm]{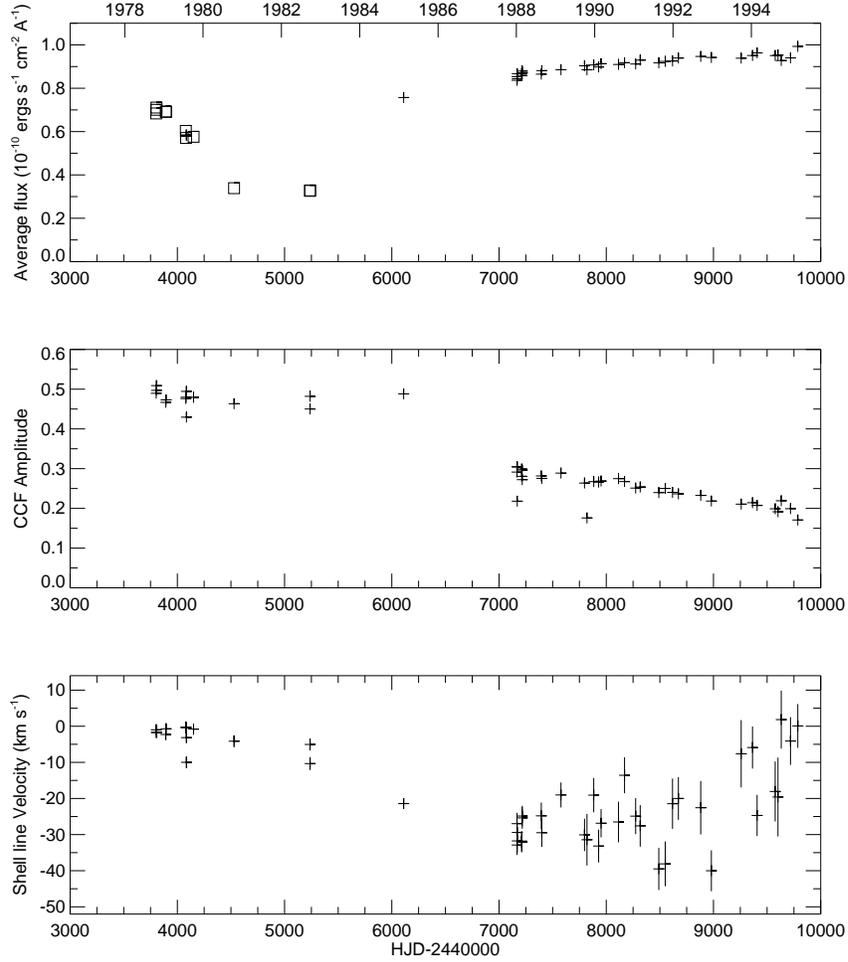}
 \caption{The shell line variations of Pleione in UV from 1979 to 1995. \emph{Top panel}: the average flux is 
computed in the wavelength region [1160, 1450] \AA; spectra recorded in large and small apertures are denoted 
as crosses and squares, respectively. \emph{Middle panel}: cross-correlation strength between the spectra of 
Pleione and supergiant star Deneb. \emph{Bottom panel}: absolute radial velocities from the cross-correlation 
functions of the shell lines.}
\label{Appendix Fig.1}
\end{figure} 
 
\newpage


\bibliographystyle{apj}
\bibliography{apj-jour,ms}

\end{document}